\documentclass[osajnl,twocolumn,showpacs,superscriptaddress,10pt]{revtex4-1}%,draft
\usepackage{amsmath,amssymb,graphicx}

\begin{document}

\title{Preparing arbitrary pure states of spatial qudits with a single phase-only spatial light modulator}
%\title{Spatial qudits accurately prepared with a single phase-only spatial light modulator}

\author{M. A. Sol\'is-Prosser}
\email{Corresponding author: msolisp@udec.cl}
\affiliation{Center for Optics and Photonics, Universidad de Concepci\'on, Casilla 4016, Concepci\'on, Chile}
\affiliation{Departamento de F\'isica, MSI-Nucleus on Advanced Optics, Universidad de Concepci\'on, Concepci\'on, Chile}

\author{$\!$A. Arias}
\affiliation{Center for Optics and Photonics, Universidad de Concepci\'on, Casilla 4016, Concepci\'on, Chile}
\affiliation{Center for Research, Instituto Tecnol\'ogico Metropolitano, A.A 54954, Medell\'in, Colombia}

\author{J. J. M. Varga}
\affiliation{Departamento de F\'isica, FCEyN, Universidad de Buenos
Aires, Buenos Aires 1428, Argentina}

\author{L.~Reb\'on}
\affiliation{Instituto de F\'isica de La Plata, Universidad Nacional
de La Plata, C.C. 67, 1900 La Plata, Argentina}

\author{S. Ledesma}
\author{C. Iemmi}
\affiliation{Departamento de F\'isica, FCEyN, Universidad de Buenos Aires, Buenos Aires 1428, Argentina}

\author{$\!$L. Neves}
\affiliation{Departamento de F\'isica, Universidade Federal de Minas Gerais,  Belo Horizonte,  MG 30123-970, Brazil}

\begin{abstract}
Spatial qudits are $D$-dimensional ($D\geqslant 2$) quantum systems carrying information encoded in the discretized transverse momentum and position of single photons. We present a proof of principle demonstration of a method for preparing arbitrary pure states of such systems by using a single phase-only spatial light modulator (SLM). The method relies on the encoding of the complex transmission function corresponding to a given spatial qudit state, onto a preset diffraction order of a phase-only grating function addressed at the SLM. Fidelities of preparation above $94\%$ were obtained with this method which is  simpler, less costly and more efficient than those that require two SLMs for the same purpose. 
\emph{\hspace{.85cm}OCIS codes:} 270.5585, 230.6120, 050.1950
\end{abstract}

%\ocis{270.5585, 230.6120, 050.1950}
%\ocis{(270.5585) Quantum information and processing; (230.6120) Spatial light modulators; (050.1950) Diffraction gratings}

\maketitle

Quantum systems are the information carriers in quantum information processing and computing protocols. While qubits (two-level systems) are the usual and most basic systems to carry out such tasks, qudits [$D$-level systems ($D>2$)] have been attracting growing interest due to their greater potential for those applications \cite{BarnettBook}. In particular, single photons are the natural choice for communications since they are easily transportable and have several degrees of freedom (DOFs) to encode information. Often, photonic qubits are encoded in the polarization, but spatial DOFs such as orbital angular momentum \cite{Mair01}, longitudinal \cite{Rossi09}, and transverse momentum-position \cite{Neves05,Hale05} are also suitable to encode qudits of higher dimensions. In the simplest approach, the latter encoding is achieved by discretizing the one-dimensional transverse modes of the photons when they are made to pass through an aperture with $D$ slits which sets the dimension of these so-called \emph{spatial qudits} \cite{Neves05}. 

Spatial qudits are relatively simple to generate and offer the possibility of working in high dimensions without cumbersome optical setups. Recently, this encoding has drawn interest for miscellaneous applications such as quantum information protocols \cite{Prosser11}, quantum games \cite{Kolenderski12}, and quantum key distribution \cite{Etcheverry13}. Therefore, the ability to prepare arbitrary pure states of such systems represents an important step toward realizations of quantum optics experiments based on it. Static amplitude and phase masks could be used for this purpose, but in practice it would be extremely difficult and time-consuming as each state intended to be prepared would require one to setup the corresponding masks in the apparatus. Programmable spatial light modulators (SLMs) can dramatically simplify this process, allowing real time manipulations of the state coefficients without physically aligning any optical components. In this regard, Ref.~\cite{Lima11} shows that by imaging the output beam of an amplitude-only SLM onto a phase-only SLM allows one to get complete and independent control of the amplitude and phase of the complex coefficients that define the qudit state. However, the use of two SLMs, besides being more costly, entails two drawbacks: (i) the overall diffraction efficiency is very low,  and (ii) in order to avoid even more losses, the image of the first SLM must match at the second one pixel by pixel, which is difficult in practice.

In this Letter, we present a proof of principle demonstration of a method for preparing arbitrary pure states of spatial qudits with a \emph{single} phase-only SLM, which has a much higher diffraction efficiency and does not require imaging systems. Among the alternative methods to represent a complex function in a single SLM \cite{Gerchberg72,Putten08} we choose, for simplicity, a technique based on Refs.~\cite{Davis99,Bagnoud04} where the amplitude information is encoded in a phase-only filter. On one hand, the required amplitude is achieved by programming a phase grating, with an appropriate modulation into the slit regions. On the other hand, the required phase value is obtained by adding a constant phase value to the phase grating. Then, the first diffraction order is selected at the Fourier plane and the filtered information is antitransformed on the final plane. The result is a light distribution corresponding to an image of the slits containing the complete complex modulation. Similar considerations can be made if the zero order were employed. However, in this case, the dead zones in the liquid crystal panel and the phase fluctuations make that unwanted light goes to that order affecting the intensity distribution \cite{Lizana08}. 

The generation of pure states of spatial qudits can be understood as follows. Let $\int\!d\mathbf{x}\,\psi(\mathbf{x})|1\mathbf{x}\rangle$ be the quantum state of a paraxial and monochromatic single-photon field, where $\mathbf{x}=(x,y)$ is the transverse position coordinate and $\psi(\mathbf{x})$ is the normalized transverse probability amplitude. When this photon is transmitted through an aperture described by a complex transmission function $T(\mathbf{x})$, its state is transformed as $\int\!d\mathbf{x}\,\psi(\mathbf{x})T(\mathbf{x})|1\mathbf{x}\rangle$.
Now, let us consider that $T(\mathbf{x})$ is an array of $D\geqslant 2$ rectangular slits of width $2a$, period $d$ and length $L(\gg a,d)$, where each slit, $\ell$, has a transmission amplitude $\beta_\ell$, i.e., $T(\mathbf{x})= {\rm rect}\!\left(x/L\right)\times\sum_{\ell=0}^{D-1}\beta_\ell\;{\rm rect}\!\left[(y-\eta_\ell d)/2a\right],$ where $\eta_\ell=\ell+(D-1)/2$. Without loss of generality and for simplicity, we will assume that $\psi(\mathbf{x})$ is constant across the region of the slits. Hence, the state of the transmitted photon will be \cite{Neves05}
\begin{equation}     \label{eq:spatial_qudit}
|\psi\rangle=\sum_{\ell=0}^{D-1}\tilde{\beta}_\ell|\ell\rangle,
\end{equation}
where $\tilde{\beta}_\ell=\beta_\ell/\sqrt{\sum_{j=0}^{D-1}|\beta_j|^2}$ and $|\ell\rangle$ denotes the state of the photon passing through the slit $\ell$.

%%%%%%%%%%%%%%%%%%%%%%%%%% FIG. MASKS %%%%%%%%%%%%%%%%%%%%%%%%%%%%%%%%%%%%%%%
\begin{figure}[b]%[htbp!]
%\centerline{\includegraphics[width=1\columnwidth]{FIG1.pdf}}
\centerline{\includegraphics[width=1\columnwidth]{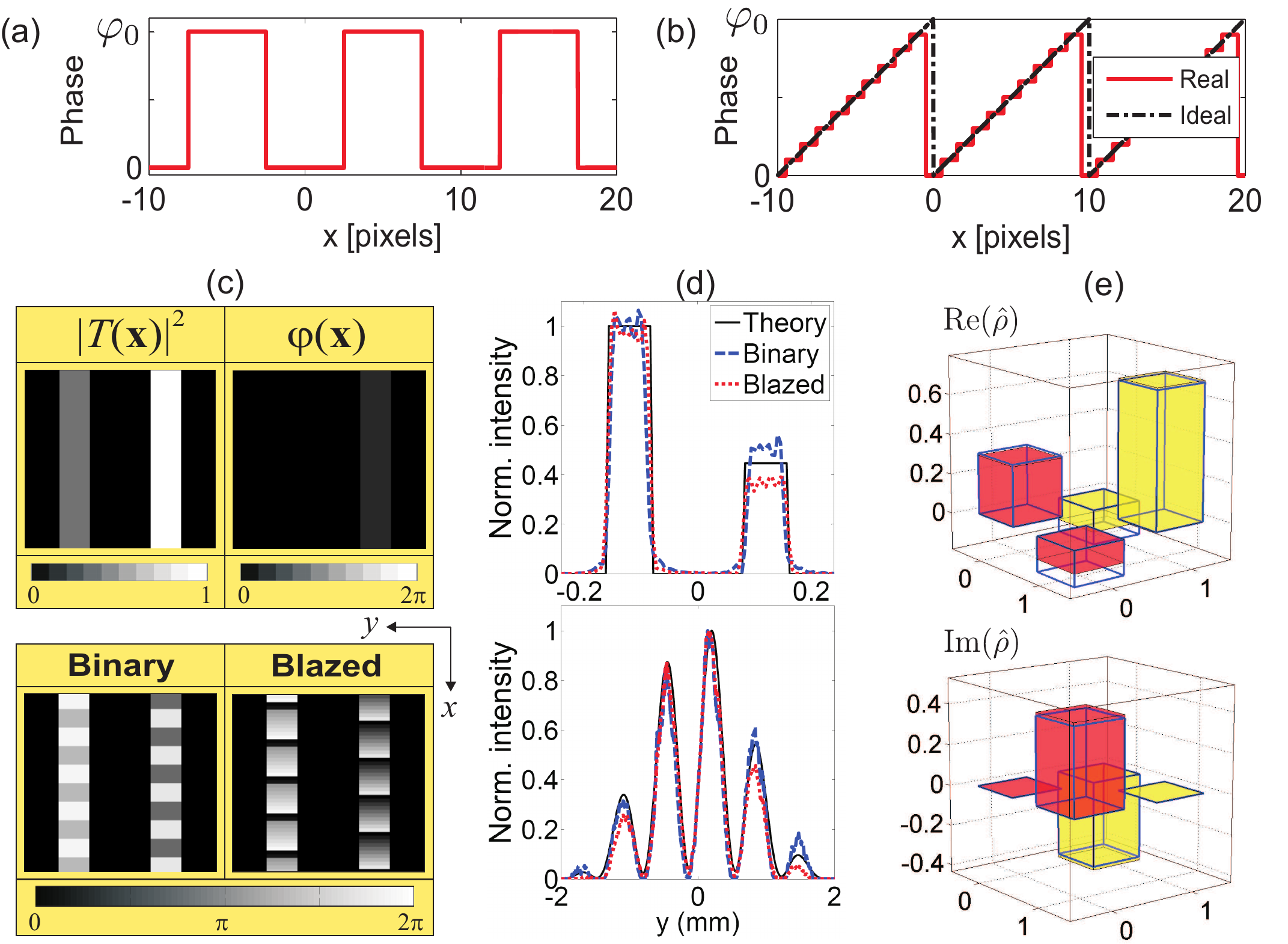}}
\caption{\label{fig:masks} (Color online). Ten-pixels period gratings: (a) binary and (b) blazed. (c) (Upper panels) amplitude and phase components of the aperture $T(\mathbf{x})$ for preparing the state
$0.56|0\rangle+0.83e^{i0.63\pi}|1\rangle$. (Lower panels) corresponding phase masks for preparing this state at the first diffraction order. (d) Normalized intensity distributions and theoretical
prediction for measurements of this state at near (upper panel) and
far field (lower panel). (e) Real and imaginary parts of the experimentally reconstructed density matrix using the blazed grating. The ideal density matrix is shown as a blue wire grid and the obtained fidelity is $F=0.996$.}
\end{figure}
%%%%%%%%%%%%%%%%%%%%%%%%%%%%%%%%%%%%%%%%%%%%%%%%%%%%%%%%%%%%%%%%%%%%%%%%%%%%%

In order to prepare arbitrary states of the form (\ref{eq:spatial_qudit}) with a single phase-only SLM we proceed as follows. A phase-only one-dimensional diffraction grating is displayed on the different regions of the SLM, each of them corresponding to a particular slit. Here, we have used two different phase profiles, binary and blazed gratings. The corresponding graphs are shown in the Figs.~\ref{fig:masks}(a) and \ref{fig:masks}(b), respectively. The efficiency of the first diffracted orders for a binary grating can be derived by using basic concepts of Fourier optics as \cite{GoodmanBook}
\begin{equation}
\epsilon_1^{\mathrm{bin}} = 
%2(1-\cos\varphi_0)/\pi^2,\label{binary-grating}
\frac{2}{\pi^2}(1-\cos\varphi_0),\label{binary-grating}
\end{equation}
where $\varphi_0$ is the phase modulation depth of the grating. Therefore, the light intensity can be modified by selecting the phase modulation $\varphi_0$. For instance, we can get an efficiency which goes from 0\% to a maximum of 40\% when $\varphi_0$ varies from 0 to $\pi$  \cite{Zhang94}. It should be mentioned that it is very simple to represent a binary grating in a SLM because only two gray levels are required. However, the maximum diffraction efficiency is limited. This can be overcome by using a blazed grating as we will discuss. In the case of blazed gratings the efficiency of the first order can be written as \cite{GoodmanBook}
\begin{equation}
\epsilon_1^{\mathrm{bla}} = 
%{\rm sinc}^2\left(1-\varphi_0/2\pi\right),\label{blazed-grating}
{\rm sinc}^2\left(1-\frac{\varphi_0}{2\pi}\right),\label{blazed-grating}
\end{equation}
where, again, $\varphi_0$ is the phase modulation depth and ${\rm sinc}(x)=\sin(\pi x)/(\pi x)$. Obviously when $\varphi_0=2\pi$ the first order efficiency has a maximum value of 100\%.  This is the usual condition for spectroscopic applications. By selecting other value for $\varphi_0$, it is possible to modulate the amount of light diffracted on the order and consequently the real amplitude of each slit. Note that Eq.~(\ref{blazed-grating}) corresponds to the ideal blazed grating with continuous modulation. Nevertheless, given that the representation of a grating period is performed through a finite number of pixels, this imposes a discretization in the phase levels for the blazed profile. Thus, it will be represented by steps as shown in Fig.~\ref{fig:masks}(b). Here, we have selected 10 quantization levels which gives us an efficiency of $\sim 97\%$ for the first order.

Independently of the selected modulation, we assign the maximum efficiency value to the maximum amplitude of the slit coefficients, i.e., $|\beta_\ell|=1$ corresponds to $\varphi_0=\pi$ for the binary modulation and $\varphi_0=(N-1)/N\times 2\pi$ for the blazed modulation, where $N$ is the number of quantization levels. Other amplitude values correspond to other values of $\varphi_0$ which are obtained from Eqs.~(\ref{binary-grating})  and (\ref{blazed-grating}). Finally, the phase of the complex coefficient in (\ref{eq:spatial_qudit}), ${\arg(\tilde{\beta}_\ell)}$, is fixed by adding a constant phase value to the phase grating. In order to avoid the introduction of additional phases in the encoding process, the phase gratings should be designed with zero mean value. However, as the SLM can only display positive phase values, the gratings are generated with a mean value equal to the half of the maximum phase modulation depth, for each kind of grating: $\pi/2$ for the binary grating and $(N-1)/N\times\pi$ for the blazed grating.

As an illustrative example, let us consider $T(\mathbf{x})$ with $D=2$ and $(\beta_0,\beta_1)=(0.67,e^{i0.63\pi})$. This aperture, whose corresponding amplitude and phase components are shown in the upper panel  of Fig.~\ref{fig:masks}(c), prepares the spatial qubit  state $0.56|0\rangle+0.83e^{i0.63\pi}|1\rangle$. In the lower panel of Fig.~\ref{fig:masks}(c) it is shown, in gray levels, the required phase distributions that are necessary to represent this particular state for each grating modulation.

%%%%%%%%%%%%%%%%%%%%%%%%%% FIG. SETUP %%%%%%%%%%%%%%%%%%%%%%%%%%%%%%%%%%%%%%%%
\begin{figure}[b]
%\centerline{\includegraphics[width=1\columnwidth]{FIG2.pdf}}
\centerline{\includegraphics[width=1\columnwidth]{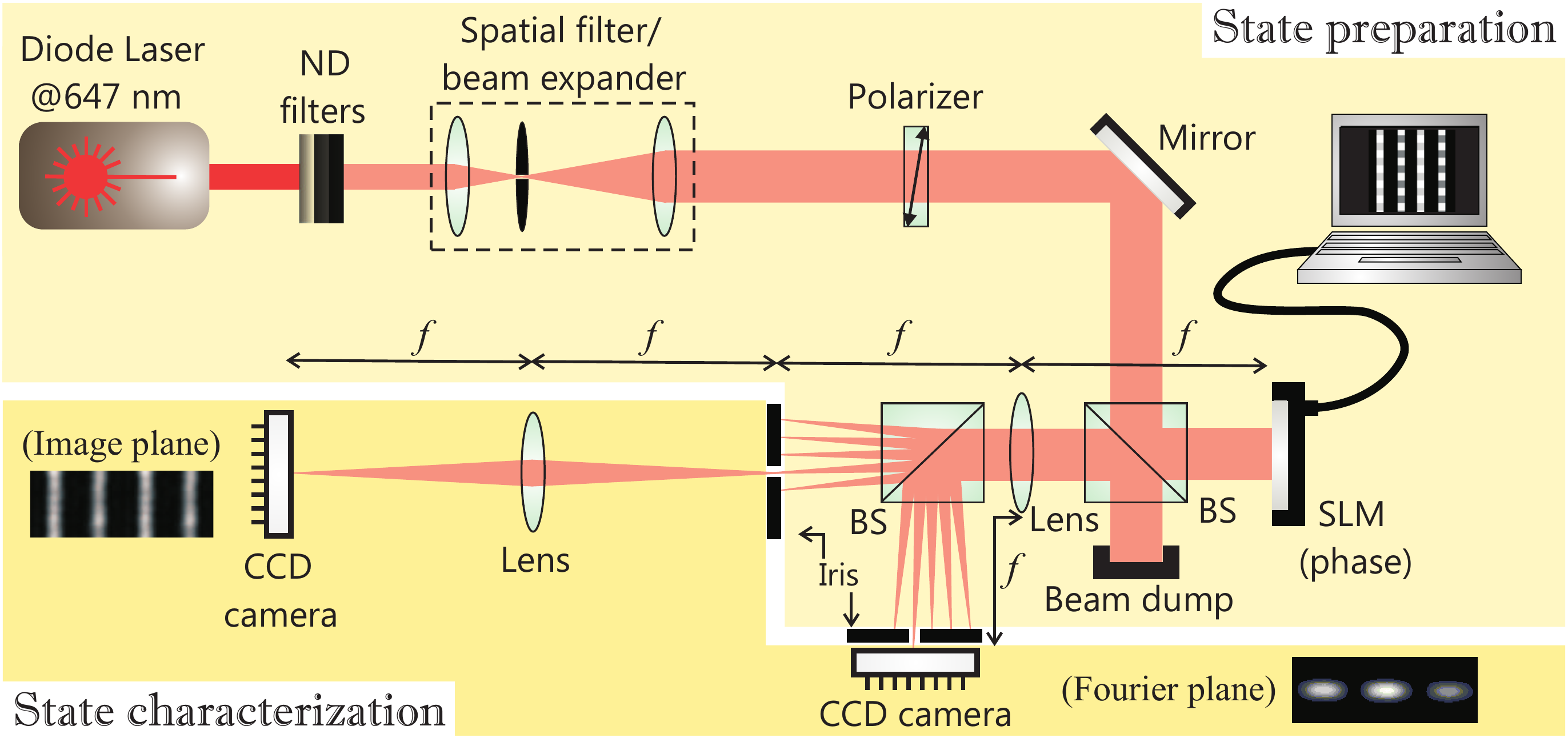}}
%\centerline{\includegraphics[width=1\columnwidth]{Neves_195601_fig2.pdf}}
\caption{(Color online) Experimental setup (details in text).}
\label{fig:setup}
\end{figure}
%%%%%%%%%%%%%%%%%%%%%%%%%%%%%%%%%%%%%%%%%%%%%%%%%%%%%%%%%%%%%%%%%%%%%%%%%%%%%

Figure~\ref{fig:setup} shows the experimental setup employed for implementing the method described above. We used a cw 647 nm single mode laser diode whose transverse spatial profile is proportional to the transverse probability amplitude of a single-photon field. In the first part of the setup used for state preparation, the attenuated laser beam was spatially filtered and collimated. Thus, the beam transverse profile, which normally impinged the  SLM, was approximately a plane wave with constant phase across the region where the slits were displayed.  The required pure phase modulation was provided by the reflective SLM (Holoeye PLUTO) and an input polarization control.  The modulated beam was split in two arms by the second beam splitter (BS). One of them allowed to obtain an image of the slits meanwhile the other was used to obtain the corresponding far field distribution. On each arm an iris diaphragm was placed at the focal plane of the transforming lens ($f=30$~cm) in order to filter the first diffracted order which carried the required information. In the second part of the setup used to perform the characterization of the states, intensity measurements were carried out with charge-coupled device (CCD) cameras. In order to register the far field distribution, the CCD of the reflected beam was placed in the Fourier plane of the slits. The light distribution of the transmitted beam was imaged onto the other CCD by means of an antitransforming lens, which avoids unwanted spurious phases \cite{GoodmanBook}. 

As mentioned earlier, for both binary and blazed gratings we used a 10-pixels period. This value provides high diffraction efficiency and enables the first diffraction order to be situated far away from the zeroth order on the Fourier plane. In this way it is easily filtered and the light distribution is not corrupted by unwanted noise. To quantify the quality of the preparation we used the fidelity, $F\equiv\langle\psi|\hat{\rho}|\psi\rangle$, between the state intended to be prepared, $|\psi\rangle$, and the density matrix of the state actually prepared, $\hat{\rho}$, which is reconstructed by tomography. Ideally, it is desirable to have $F=1$. The tomographic process is performed by measurements  at transverse positions corresponding to the  states $\{|0\rangle,\ldots,|D-1\rangle\}$ in the near field  and $\left\{|\chi\rangle_\xi=\sum_{\ell=0}^{D-1}e^{i\ell\xi}|\ell\rangle/\sqrt{D}\right\}$ in the far field, where the set $\{\xi\}$ depends on $D$; for arbitrary qubits $\{\xi=j\pi/2\}_{j=0}^{3}$ \cite{Lima08}; for pure qudits only, one can numerically find the phases of a state whose interference pattern is the closest to the experimental one. We have measured the intensities at those positions and applied maximum likelihood technique to the recorded data in order obtain the best density matrix estimation consistent with the requirements of a physical state \cite{Fiurasek01}. %\cite{James01}. 

%%%%%%%%%%%%%%%%%%%%%%%%%%%%%%%%% FIG. BLOCH %%%%%%%%%%%%%%%%%%%%%%%%%%%%%%
\begin{figure}%[htbp]
%\centerline{\includegraphics[width=1\columnwidth]{FIG3.pdf}}
%\centerline{\rotatebox{-90}{\includegraphics[width=.76\columnwidth]{Neves_195601_fig3.pdf}}}
%\centerline{\includegraphics[height=.35\textheight,width=.85\columnwidth]{Neves_195601_fig3.pdf}}
\centerline{\includegraphics[width=0.98\columnwidth]{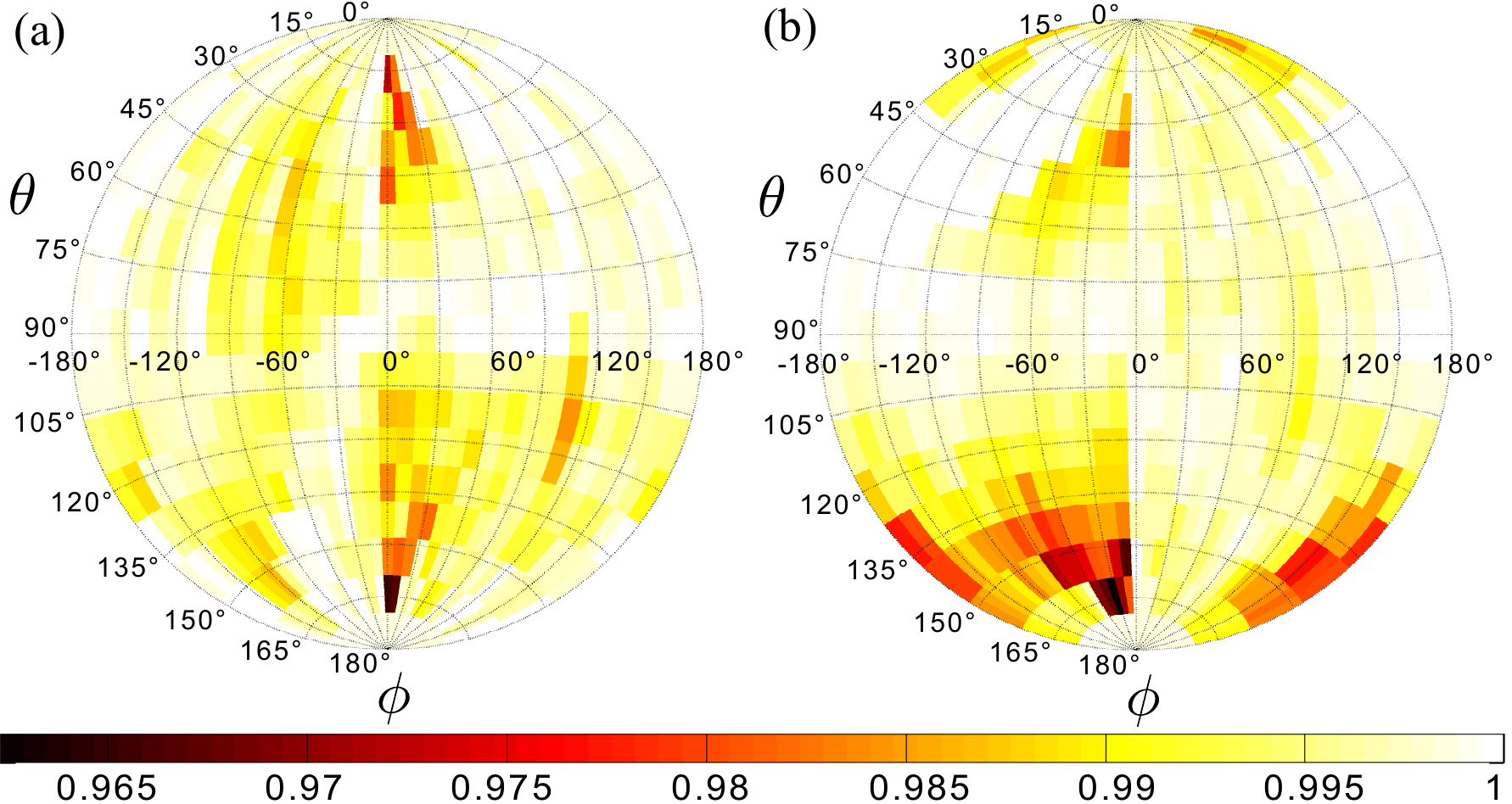}}
%\vspace{-2.3cm}
%\vspace{-1.8cm}
\caption{(Color online) Bloch spheres showing the fidelities of preparation of  spatial qubit states using (a) binary and (b) blazed gratings.
The latitude $\theta\in[0,\pi]$ and the longitude $\phi\in(-\pi,\pi]$ parametrizes an arbitrary pure state $|\psi\rangle = \cos(\theta/2)|0\rangle
+ e^{i\phi}\sin(\theta/2)|1\rangle$ on the sphere surface.}
\label{fig:Bloch}
\end{figure}
%%%%%%%%%%%%%%%%%%%%%%%%%%%%%%%%%%%%%%%%%%%%%%%%%%%%%%%%%%%%%%%%%%%%%%%%%%

\begin{figure}[t]
%\centerline{\includegraphics[width=1\columnwidth]{FIG4.pdf}}
\centerline{\includegraphics[width=1\columnwidth]{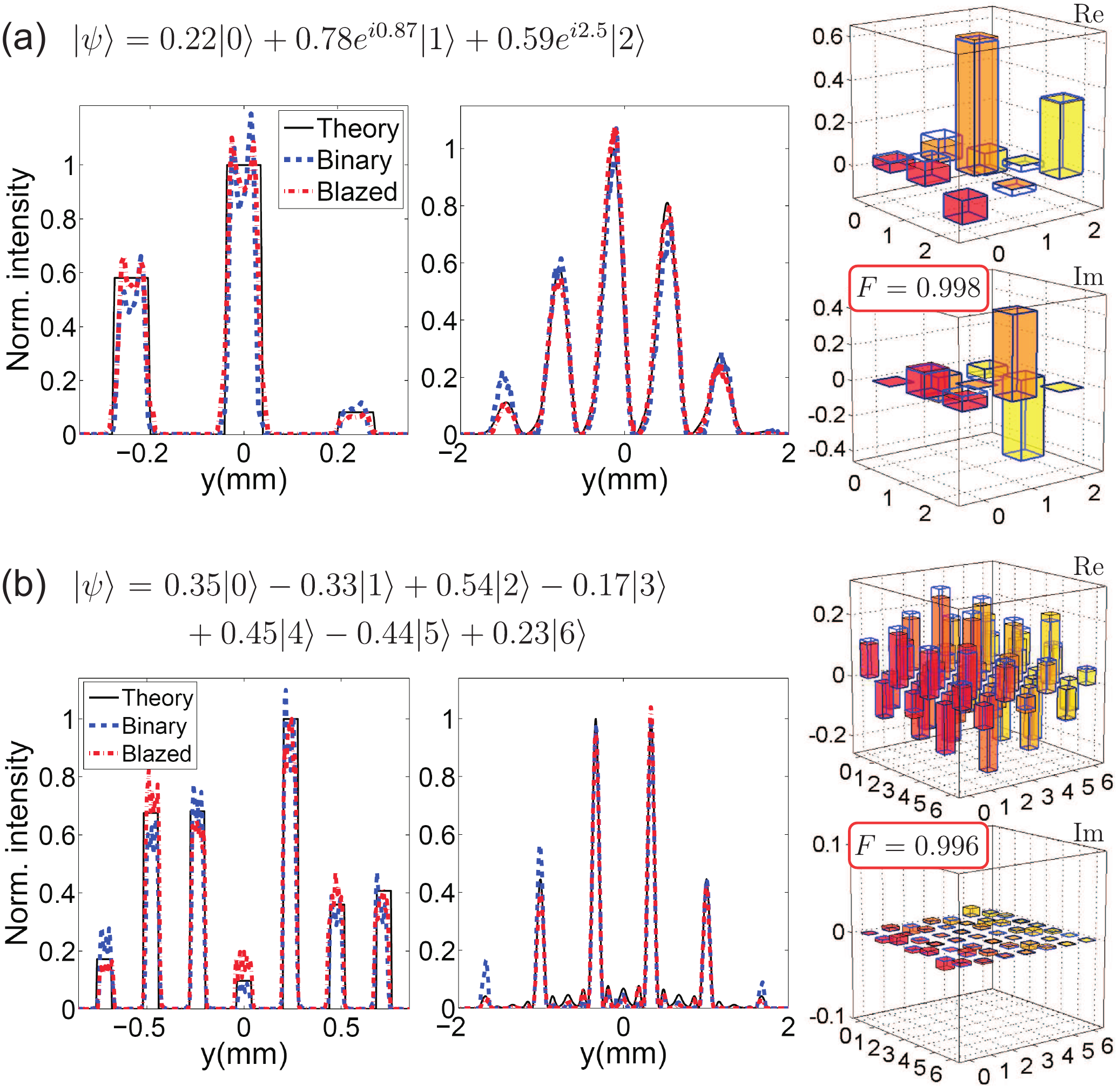}}
\caption{(Color online) Three- (a) and seven-dimensional (b) qudit states with their corresponding normalized intensity distributions for measurements at near and far field, and the experimentally reconstructed density matrices using the blazed grating. The ideal density matrices are shown as blue wire grids and the obtained fidelities are shown in the insets.}
\label{fig:qudits}
\end{figure}

For spatial qubits, Figs.~\ref{fig:masks}(d) and \ref{fig:masks}(e) show, as an example for the particular state discussed earlier, the measured intensity distributions in the near and far fields, and the corresponding tomography, respectively. In Fig.~\ref{fig:Bloch}, each Bloch sphere shows the fidelities of preparation for 561 states uniformly distributed on the surface, using either binary [Fig.~\ref{fig:Bloch}(a)] or blazed [Fig.~\ref{fig:Bloch}(b)] grating. In both cases, we obtained fidelities above $96\%$ and average fidelities of $99.6\%$ as shown in Table~\ref{table:mean_fids}. The discontinuity that appears around the meridian $\phi=0$ is, possibly, due to the different phase fluctuations at each gray level. For instance, although 0 and $2\pi$ are geometrically equivalent, the applied voltages are different and the fluctuations in one case are higher than in the other \cite{Lizana08} (We will discuss this problem in more detail elsewhere). The loss of fidelity observed around the poles could be attributed to a slight misalignment of the laser beam with respect to the slits center. For spatial qudits of different dimensions, hundreds of states have been prepared and the fidelities were above $94\%$. Figure~\ref{fig:qudits} shows two samples and Table~\ref{table:mean_fids} shows the number of states prepared per type of grating and per qudit dimension with their corresponding average fidelities. Regarding the type of grating, either binary or blazed, the overall quality of the prepared states is very similar. However, as discussed earlier, blazed gratings provide better diffraction efficiency than binary ones. This enhanced efficiency can be important to increase the signal-to-noise ratio when working with single-photon sources. 

%%%%%%%%%%%%%%%%%%%%%%%%%%% TABLE - FFIDELITIES %%%%%%%%%%%%%%%%%%%%%%%%%%%%%%%%
\begin{table}[t]
  \caption{Average fidelities of preparation for $D$-level qudits.}
  \begin{center}
    \begin{tabular}{ccccccccccc}
    \hline
$D$ && 2 && 3 && 4 && 5 && 7  \\
    \hline
$\#\ {\rm of\ states}/{\rm grating}$ && 561 && 24 && 70 && 25 && 94  \\
$ \bar{F}_{\rm bin}$ && 0.996 && 0.995 && 0.985 && 0.968 && 0.970  \\
 $\bar{F}_{\rm bla}$ && 0.996 && 0.996 && 0.991 && 0.971 && 0.977   \\
    \hline
    \end{tabular}
  \end{center}
\label{table:mean_fids}
\end{table}
%%%%%%%%%%%%%%%%%%%%%%%%%%%%%%%%%%%%%%%%%%%%%%%%%%%%%%%%%%%%%%%%%%%%%%%%%%%%%%

We have compared the luminous efficiency of our setup with the one that employed two SLMs \cite{Lima11}. In case of using a blazed grating, our scheme is $1/\eta$ times more efficient, where $\eta$ represents the light attenuation due to polarizing elements needed to obtain pure amplitude modulation with a twisted nematic SLM \cite{Marquez01,Moreno04}. The factor $\eta$ depends on the parameters involved in each particular case (wavelength, twist angle, birefringence of the SLM, etc.), but a typical value is about 0.1.

In conclusion, we have shown that a single phase-only SLM enables the preparation of arbitrary pure states of spatial qudits. Fidelities of preparation above $94\%$ were obtained and a higher efficiency respect to other methods is expected. It is important to stress that this method can be useful not only for preparation but also for assisting the measurement of such systems. It enables the implementation of the strategy proposed in Ref.~\cite{Prosser10} by projecting the image of the state to be measured at the SLM and addressing at this device the phase grating function corresponding to the measurement state. 
A detection at the center of the filtered diffraction order at the Fourier plane will accomplish the process allowing one to obtain the statistics of arbitrary observables. Therefore, the method presented here may become a valuable tool for experiments based on this encoding.

%%%%%%%%%%%%%%%%%%%%%%%%%%%%%%%%% QUDITS %%%%%%%%%%%%%%%%%%%%%%%%%%%%%%%%%
%%%%%%%%%%%%%%%%%%%%%%%%%%%%%%%%%%%%%%%%%%%%%%%%%%%%%%%%%%%%%%%%%%%%%%%%%%

%\begin{acknowledgments}
We thank P. Kolenderski and A. Calatayud for helpful discussions.
This work was supported by CONICYT PFB08-24 and Milenio ICM P10-030-F (Chile); UBACyT 20020100100689, CONICET PIP 112-200801-03047 and ANPCYT PICT 2010-02179 (Argentina). M.A.S.P. acknowledges financial support from CONICYT. 
%\end{acknowledgments}

\end{document}